\documentclass[runningheads]{svmult}

\input psfig.sty

\usepackage{makeidx}   % allows index generation
\usepackage{graphicx}  % standard LaTeX graphics tool
\usepackage{subeqnar}  % subnumbers individual equations
\usepackage{multicol}  % used for the two-column index
\usepackage{physprbb}  % modified textarea for proceedings,
\makeindex             % used for the subject index
\def\ltsima{$\; \buildrel < \over \sim \;$}
\def\lsim{\lower.5ex\hbox{\ltsima}}
\def\gtsima{$\; \buildrel > \over \sim \;$}
\def\gsim{\lower.5ex\hbox{\gtsima}}

\begin{document}
\title*{The Masses of Lyman Break Galaxies}
\toctitle{The Masses of Lyman Break Galaxies}
\titlerunning{The Masses of Lyman Break Galaxies}
\author{Joel R. Primack\inst{1}
\and Risa H. Wechsler\inst{2}
\and Rachel S. Somerville\inst{3}}
\authorrunning{J.R. Primack, R.H. Wechsler, and R.S. Somerville}

\institute{Physics Department, University of California, Santa Cruz,
CA 95064 USA 
\and Department of Physics, University of Michigan, Ann Arbor, 
MI 48109 USA
\and Department of Astronomy, University of Michigan, Ann Arbor, 
MI 48109 USA}

\maketitle              % typesets the title of the contribution

\begin{abstract}
Data on galaxies at high redshift, identified by the Lyman-break
photometric technique, can teach us about how galaxies
form and evolve.  The stellar masses and other properties of such
Lyman break galaxies (LBGs) depend sensitively on the details of star
formation.  In this paper we consider three different star formation
prescriptions, and use semi-analytic methods applied to the
now-standard $\Lambda$CDM theory of hierarchical structure formation
to show how these assumptions about star formation affect the
predicted masses of the stars in these galaxies and the masses of the
dark matter halos that host them.  We find that, within the rather
large uncertainties, recent estimates of the stellar masses of LBGs
from multi-color photometry are consistent with the predictions of all
three models.  However, the estimated stellar masses are more
consistent with the predictions of two of the models in which star
formation is accelerated at high redshifts $z\gsim3$, and of these
models the one in which many of the LBGs are merger-driven starbursts
is also more consistent with indications that many high redshift
galaxies are gas rich.  The clustering properties of LBGs have put
some constraints on the masses of their host halos, but due to
similarities in the halo occupation of the three models we consider
and degeneracies between model parameters, current constraints are not
yet sufficient to distinguish between realistic models.
\end{abstract}

\section{Introduction}
A great deal of effort devoted to determining the cosmological
parameters has recently paid off.  But, although there is good
evidence that the cosmological parameters are roughly
$\Omega_m=0.3$, $\Omega_\Lambda=0.7$, and $h=0.7$, 
and that $\Lambda$CDM with these parameters is a good fit to the
observed universe \cite{isss1}, this theory does not make unique
predictions regarding the masses and other properties of galaxies at
high redshift.  Galaxy properties in cosmological theories also depend
on assumptions about uncertain aspects of star formation, supernova
feedback, and dust obscuration.  Here we will focus on star formation.

We consider three different models of star formation, differing in the
way that the star formation rate depends on galaxy properties, and
discuss the implications for masses, clustering, and other properties
of Lyman break galaxies (LBGs) in semi-analytic models.  These models 
\cite{spf} all assume exactly the same underlying $\Lambda$CDM
model with the parameters above, so the properties of the dark matter
halos at any given redshift and the halo merging histories are the
same.  We also make the same assumptions in each model regarding the
initial mass function (IMF), which we assume to be Salpeter between
0.1 and 100 $M_\odot$. 
We use the GISSEL00 stellar population synthesis models of Bruzual \&
Charlot \cite{bc} with solar metallicity, and a simple model for dust
obscuration, in which the optical depth is a power-law function of the
unobscured ultraviolet luminosity:
\begin{equation}
\tau_{UV} = \tau_{UV*} (L_{UV}/L_{UV*})^\beta \; , \label{eq:dust}
\end{equation}
with $\tau_{UV*}$ an adjustable parameter and $\beta=0.5$
\cite{wangheckman}.

The modern approach to semi-analytic modeling was pioneered by White
\& Frenk \cite{wf91}, and further developed by them and their
collaborators in \cite{kwg93} and \cite{cole94}.  In \cite{sp}, we
reviewed and extended this work, and applied it to high-redshift
galaxies \cite{spf} using three simplified models of star formation.
The three star formation models we consider span the range of models
proposed for high redshift galaxy formation.

The simplest model, termed Constant Efficiency Quiescent,
assumes that the quiescent star formation rate per
unit mass of cold gas is constant:
\begin{equation}
\dot m_* = \frac{m_{cold}}{\tau_*} \qquad\qquad {\rm (CEQ)} \; .
\end{equation}
We showed in \cite{spf} that the resulting predictions of this model
are similar to those of \cite{cole94} and to more detailed treatment
\cite{baugh98,governato98} from the same group that 
included starbursts from major mergers.
An alternative Accelerated Quiescent model assumes that
\begin{equation}
\dot m_* = \frac{m_{cold}}{\tau_{dyn}} \qquad\qquad {\rm (AQ)} \; .
\end{equation}
The predictions of our AQ model are like those of \cite{kwg93} even
though those authors included starbursts from major mergers, because
the above star formation prescription (based on data on star formation
in nearby galaxies \cite{kennicutt}) converts gas to stars so
efficiently at high 
redshifts (since typically $\tau_{dyn} \propto 1/(1+z)$).  Finally, we
consider a Collisional StarBurst model in which the quiescent star
formation efficiency is constant, but where there is also a burst mode of
star formation triggered by galaxy interactions:
\begin{equation}
\dot m_* = \frac{m_{cold}}{\tau_*} + (\dot m_*)_{\rm starbursts} \quad
{\rm (CSB)} \; ,
\end{equation}
where $(\dot m_*)_{\rm starbursts}$ is due to
bursts in merging galaxies. The efficiency of star formation in these
bursts is scaled according to a model based on hydrodynamical
simulations \cite{mihoshernquist}, in which the efficiency scales as a
power-law function of the mass ratio of the merger (see \cite{spf} for
details).  We find in our CSB semi-analytic model that most of the
star formation at redshifts above unity occurs in starbursts driven by
minor mergers, in which the merging satellite has mass less than
$\frac{1}{3}$ that of the central galaxy.
As noted, our CEQ model is similar to the Durham group model of
several years ago \cite{baugh98}.  At this conference, Baugh presented
preliminary results from an alternative model that is similar to our
CSB model.

% fig 1
\begin{figure}
\noindent \begin{minipage}[t]{2.25in} \centering %\leavemode
 \psfig{file=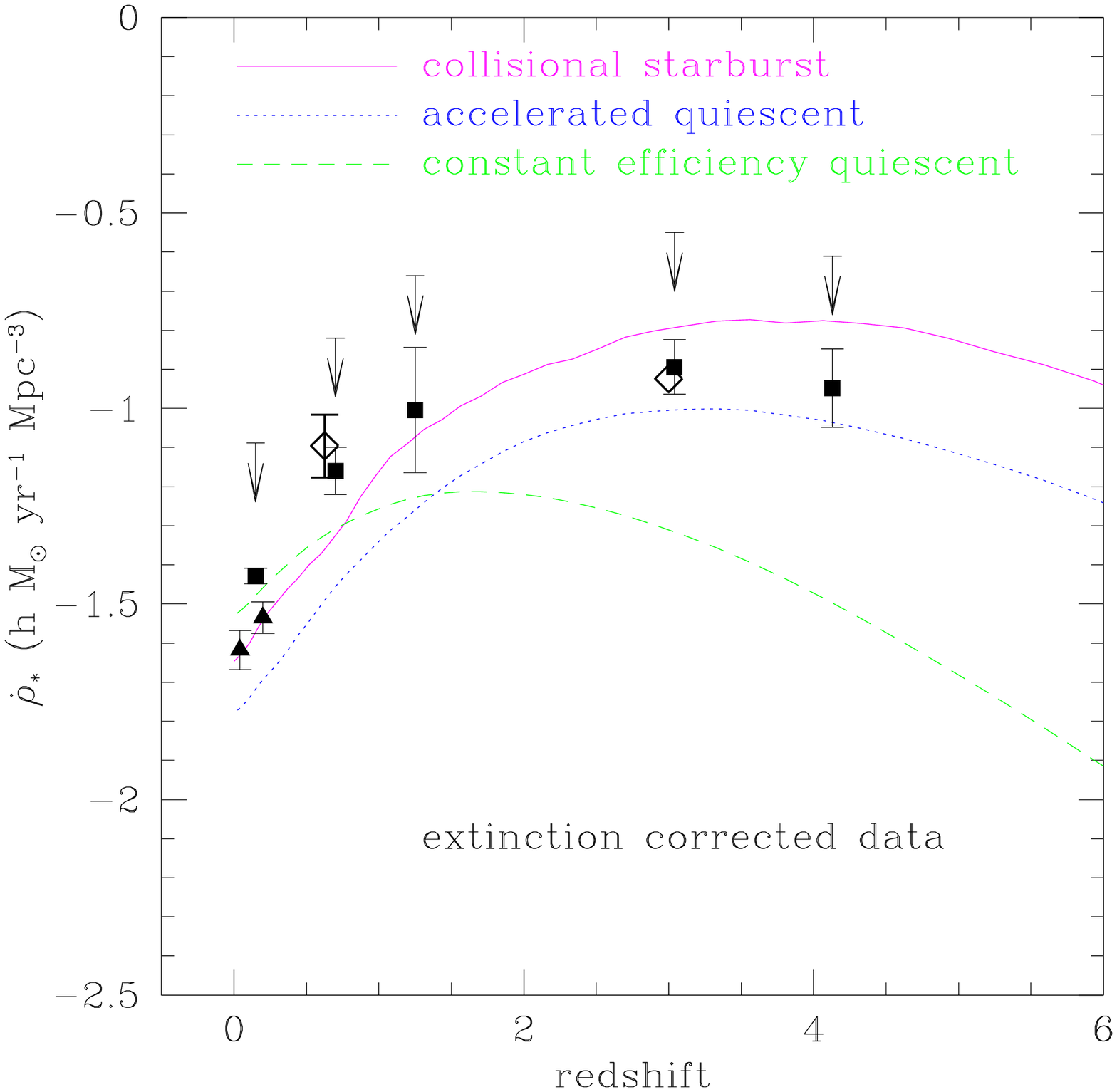,width=2.25in} \end{minipage} \hfill
\begin{minipage}[t]{2.25in} \centering %\leavemode
 \psfig{file=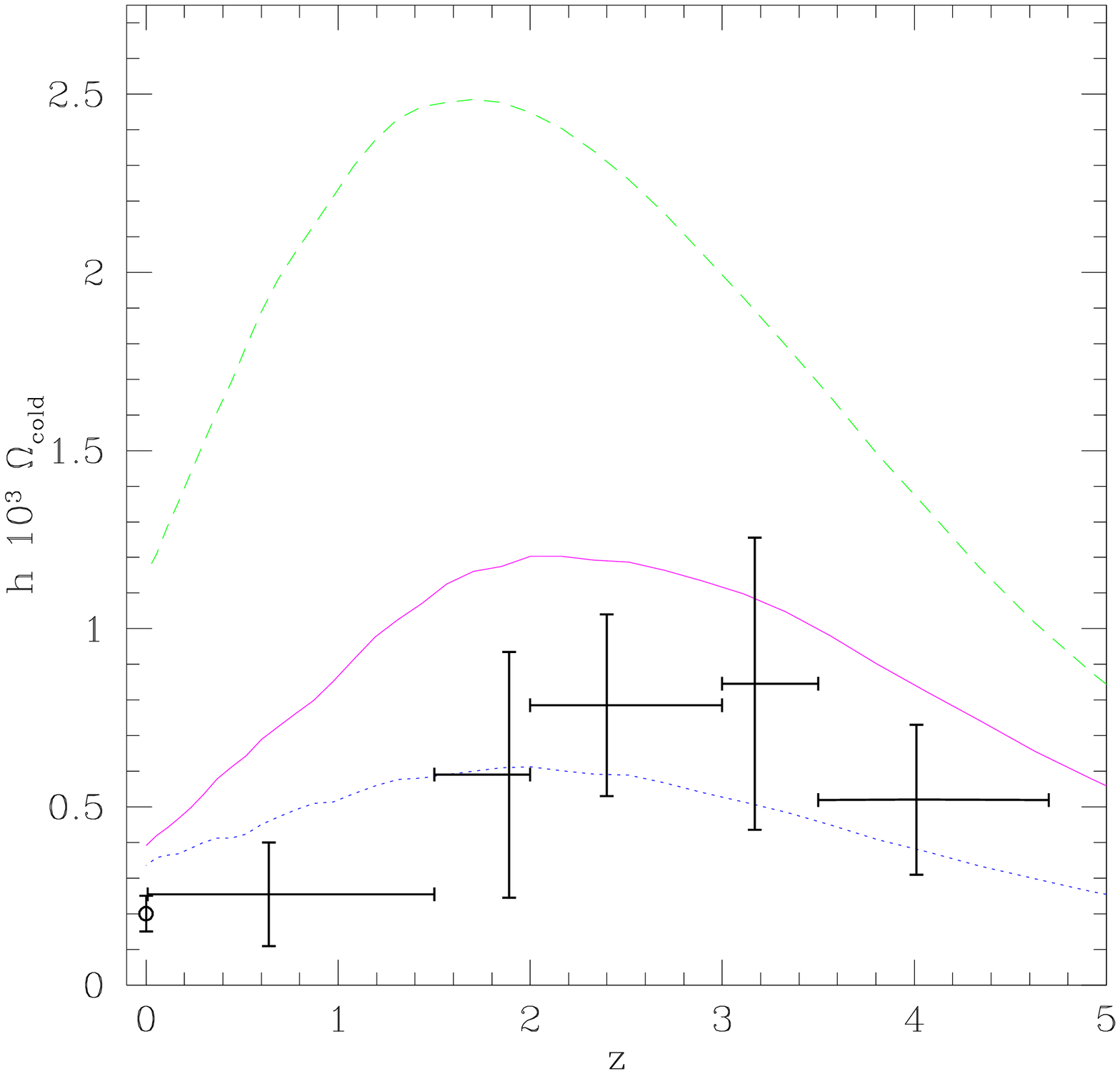,width=2.25in} \end{minipage}
\caption{Madau plot of star formation rate density (a) and cosmological density of neutral hydrogen
(b) for our three models CSB, AQ, and CEQ.  (From \protect\cite{spf}.)}
\end{figure}

Figure 1a shows the star formation rate density as a function of
redshift for these three models.  The CEQ model does not
produce as many stars at high redshifts $z\gsim3$ as the
extinction-corrected observations indicate (see \cite{spf} for
references), while both AQ and CSB models are acceptable in this
regard, indicating that star formation at high redshift must be more
efficient than locally.  This argument is reinforced by the failure of
the CEQ model to produce as many stars at $z>2$ as are indicated by
the fossil evidence (see Fig. 12 of \cite{spf}).  

However, the AQ model converts gas to stars so efficiently that it may
not have as much neutral hydrogen at $z>2$ as is indicated by the data
on damped Lyman-alpha systems --- see Fig. 1b.  The AQ model may also
not have enough gas to fuel quasars at high redshifts
\cite{kaufhaehnelt}, while the CSB model seems acceptable.

Further evidence that favors the CSB over the AQ model comes from the
predicted LBG luminosity function.  The CSB model predicts as many
bright LBGs as are observed, although it slightly overpredicts the
number of fainter ones, possibly because the dust obscuration
prescription Eq. (\ref{eq:dust}) is unrealistic in predicting very
little extinction for lower-luminosity galaxies.  But the AQ
luminosity function predicts fewer bright LBGs than observed at $z=3$,
and far fewer than observed at $z=4$ (see Figs. 4-7 of \cite{spf}).
Bright LBGs only occur in massive halos in the CEQ and AQ models, and
there are fewer such halos at higher redshifts.

\section{LBG Masses}

\begin{figure}
\centerline{\psfig{file=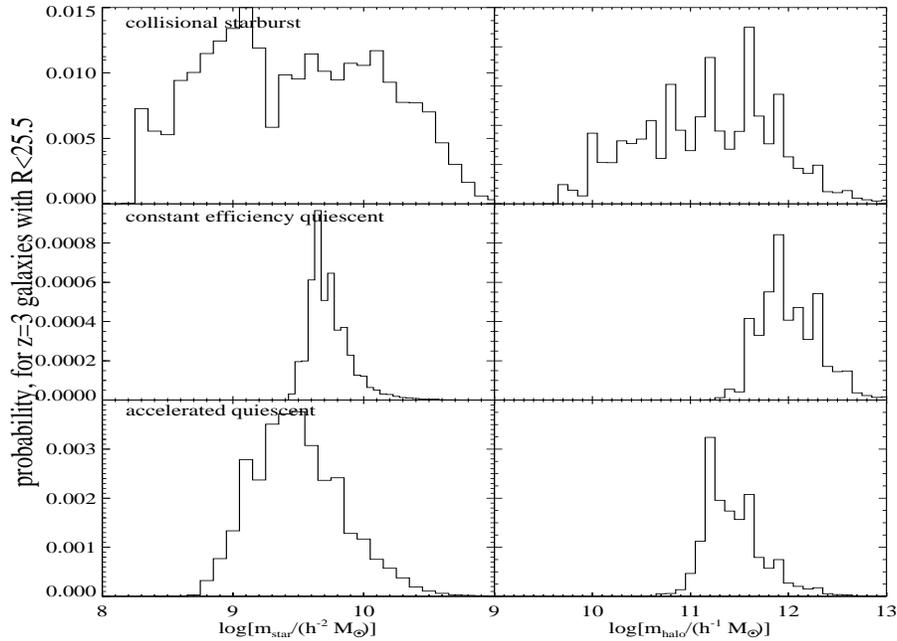,height=3.5in,width=4.9in}}
\caption{Stellar and halo masses for galaxies brighter in $R_{AB}$
than 25.5 at $z=3$.
}
\label{fig:6panel}
\end{figure}

At this conference we have seen that analysis of the Hubble Deep Field
data on LBGs \cite{papovich} indicates that their stellar masses lie
in the range $10^9 - 10^{11} M_\odot$, with a geometric mean of
$6\times 10^9 M_\odot$ (assuming a Salpeter IMF and solar metallicity,
as we have done).  Ground-based data on somewhat brighter LBGs
indicate stellar masses in the same range, with a slightly higher
median (see especially Fig. 10b of \cite{shapley}).  The predicted
ranges of stellar masses and halo masses for all three of our models
are shown in Fig. \ref{fig:6panel}.  (The results for the CSB model
are similar to those represented by the histogram in the top right
panel of Fig. 16 of \cite{spf}, except that here they for
$R_{AB}<25.5$ while there they were for $V_{606}<25.5$.)  It is
interesting that the distributions in the three models have similar
medians, but very different widths. Perhaps counter-intuitively, the
CSB model actually has \emph{more} galaxies with large stellar masses
than the other models! This is because there is more star formation
activity at $z>3$ in this model.
However, in interpreting
Fig. \ref{fig:6panel} it is important to keep in mind that
the LBGs are predicted to be systematically fainter at a given stellar
mass in the CEQ model.

% fig 2

% fig 3
\begin{figure}
\centerline{\psfig{file=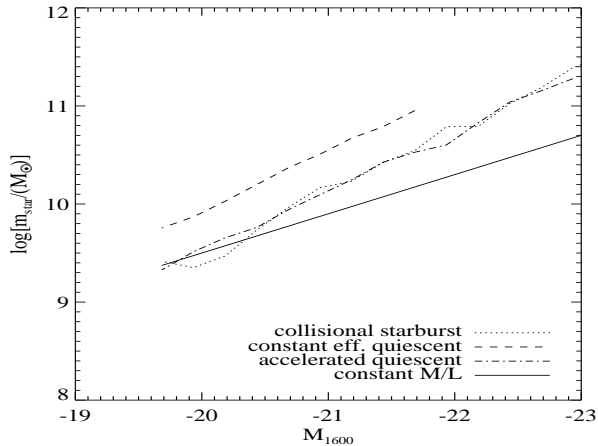,height=2.5in,width=3.5in}}
\caption{Median stellar masses vs. rest-frame UV magnitude for CEQ,
AQ, and CSB models.  This relation is sensitive to the prescription
for dust extinction; models that do not include dust extinction
are closer to constant $M/L$.}
\label{fig:mstar_mag}
\end{figure}

It may therefore be more illuminating to look at the predicted
relationship between stellar mass and rest-frame UV luminosity in all
three models, shown in Fig. \ref{fig:mstar_mag}.  The stellar masses
deduced from the HDF-N data (see the lower right panel of Fig. 17 of
\cite{papovich}) agree well in zero point and slope with the
predictions of the CSB and AQ models, but the stellar masses in the
CEQ model are higher by about a factor of 2.5.  However, the ground
based data \cite{shapley} do not show as clear a correlation of
stellar mass with UV luminosity.  Also, the deduced stellar masses are
sensitive to the IMF and dust extinction, so there are large
uncertainties making it impossible to rule out any of the models on
this basis.

\section{LBG Clustering}

In principle, another way of estimating the masses of LBGs is via
their clustering, since dark matter halos of higher mass are expected
to be more correlated (e.g. \cite{mowhite}).  However, collisions
between lower-mass halos are also more correlated than the halos
themselves, since the collisions occur preferentially in denser
regions \cite{kolatt} --- thus knowledge of the LBG host halo masses
from their clustering properties does not uniquely specify whether the
LBGs are associated with galaxies in massive subhalos or with galaxy
collisions.  When we used N-body simulations combined with
semi-analytic models to compute clustering properties of LBGs, we
\cite{risa} found that the CEQ, AQ, and CSB models all predicted
similar LBG clustering on both short and long scales, in general
agreement with the available data --- 
though CEQ produces the most clustered galaxies and is only marginally
consistent.  The similarity of LBG clustering properties in the three
models reflects the larger similarity than might be expected in the
dark matter halos that they occupy; we explored the physical reasons
for this in \cite{risa}.  However, the detailed clustering properties
are affected by model ingredients that are still quite uncertain ---
for example, the efficiency of converting gas into stars in a galaxy
collision --- so further theoretical constraints on these parameters
combined with recently improved observational constraints on LBG
clustering may improve the potential to distinguish between modes of
star formation.

A simple analytic model that constrains the dark matter halo masses
hosting LBGs using data on their number density and clustering has
shown that, if the typical number of galaxies in a halo of mass $M$ is
$N(M) = (M/M_1)^S$ for $M>M_{\rm min}$, current observational
constraints on this model indicate that LBGs occupy halos greater than
about $M \sim 10^{10} h^{-1} M_\odot$, with a power-law occupation of
$S\sim0.8$, so that typical LBGs reside in halos of a few $\times
10^{11} h^{-1} M_\odot$ \cite{bullockws}.  Such a model can also be
used to compare the host halo masses of different populations of
objects, and has been used \cite{moustakass} to show that
highly-clustered EROs inhabit halos that are an order of magnitude or
two more massive than LBGs.
The only possibly discrepant data on LBG clustering is the
suggestion that the correlation length is a strong function of LBG
brightness \cite{giavalisco}, in disagreement with each of our models
\cite{risa}. However, this interpretation is controversial 
(cf. \cite{risa,arnouts}).

\section{Conclusions}
While the masses of LBGs would seem to provide key information about
their nature, due to the considerable uncertainties in our modeling
and in deriving stellar or halo masses from the data, it is not
currently possible to rule out any of the three very different recipes
for star formation considered here on this basis.  All three models
are also roughly consistent with recent observational
estimates of LBG clustering.  However, the CEQ model predicts
systematically higher stellar masses and also far too few bright LBGs
especially at higher redshifts, and the AQ model may use up gas too
efficiently to be consistent with other data.  One way to check
whether it is really true that many of the high-redshift bright
galaxies are collision-driven starbursts is to see whether the
morphologies of these objects resemble those produced in
hydrodynamical simulations of interacting gas-rich galaxies, which are
presently underway (see \cite{racheldisks} for preliminary results).

\end{document}